\documentclass[preprint,12pt]{elsarticle}

\usepackage{epsfig}

\usepackage{amssymb}

\usepackage{color}

\newcommand{\MARKI}[1]{#1}
\newcommand{\MARKII}[1]{#1}
\newcommand{\MARKIII}[1]{#1}

\newcommand{\tetherlen}{\MARKII{L_t}}

\journal{Acta Astronautica}

\begin{document}

\begin{frontmatter}



\title{Electric sail, photonic sail and deorbiting applications of the
  freely guided photonic blade}


\author[FMI]{Pekka Janhunen\corref{cor1}}
\ead{pekka.janhunen@fmi.fi}
\ead[url]{http://www.electric-sailing.fi}

\address[FMI]{Finnish Meteorological Institute, Helsinki, Finland}
\cortext[cor1]{Corresponding author}

\begin{abstract}
We consider a freely guided photonic blade (FGPB) which is a
centrifugally stretched sheet of photonic sail membrane that can be
tilted by changing the centre of mass or by other means. The FGPB can
be installed at the tip of each main tether of an electric solar wind
sail (E-sail) so that one can actively manage the tethers to avoid
their mutual collisions and to modify the spin rate of the sail if
needed. This enables a more scalable and modular E-sail than the
baseline approach where auxiliary tethers are used for collision
avoidance. For purely photonic sail applications one can remove the
tethers and increase the size of the blades to obtain a novel variant
of the heliogyro that can have a significantly higher packing density
than the traditional heliogyro. For satellite deorbiting in low Earth
orbit (LEO) conditions, analogous designs exist where the E-sail
effect is replaced by the negative polarity plasma brake effect and
the photonic pressure by atmospheric drag. We conclude that the FGPB
appears to be an enabling technique for diverse applications. We also outline a way
of demonstrating it on ground and in LEO at low cost.
\end{abstract}

\begin{keyword}
electric sail \sep 
photonic sail \sep 
propellantless propulsion \sep
plasma brake \sep
deorbiting \sep
atmospheric drag deorbiting
\end{keyword}

\end{frontmatter}



\section{Introduction}

The spinning electric solar wind sail (E-sail)
\cite{paper1,paper2,RSIpaper} and the spinning heliogyro photon sail
\cite{MacNeal1967,Blomqvist2009} have been proposed for propellantless
interplanetary travel. Likewise, various LEO deorbiting devices based
on Coulomb drag \cite{paper3,paper8} and atmospheric drag
\cite{CubeSail,DEORBITSAIL} have been proposed for mitigating the space debris
problem by deorbiting satellites after their mission is complete and
by deorbiting already existing junk objects by attaching braking devices to them.

\begin{figure}
\centerline{\includegraphics[width=0.5\columnwidth]{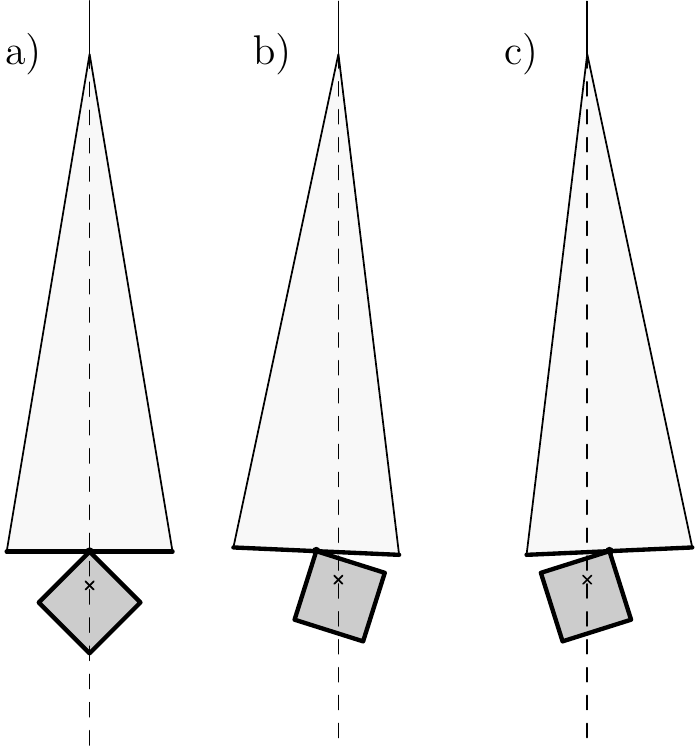}}
\caption{
Photonic blade hanging in gravity field.
Moving the ballast mass right (b) or left (c) changes the centre of
gravity (X) of the system so that a photonic torque about the vertical axis gets applied to the blade
and it starts to turn about the vertical axis.
}
\label{fig:Hang}
\end{figure}

In this paper we consider a freely guided photonic blade (FGPB) which is part of a
spinning system so that it is kept stretched by the centrifugal
force. In ground-based conditions we can mimic the centrifugal force
by gravity (Fig.~\ref{fig:Hang}a). If a ballast mass is moved sideways
(Fig.~\ref{fig:Hang}b,c), a difference between the centre of mass and
centre of photon pressure is created and consequently the photon
pressure starts to turn the blade about the vertical axis. This gives
a simple way to control the blade's orientation with respect to
sunlight and thus to control the direction of the photonic thrust
vector.

\MARKI{Throughout the paper we treat the heliogyro photonic blades as
  rigid objects. Although the blade is in reality free to bend and is
  not made of rigid material, the rigid body approximation is valid if
  the shape of the blade does not differ markedly from a planar
  surface.  This is the case if the membrane is lightweight in
  comparison to the mass of the remote unit which controls it or if
  the spin of the sail is fast so that the centrifugal force acting on
  the blade is much larger than the actuated part of the photon
  pressure force. If the blade is made very long and if it is actuated
  from either end, it tends to twist in response to actuation instead
  of tilting uniformly. Problems due to twisting and bending increase
  in severity if the blade's aspect angle $K$
  \footnote{\MARKI{See Nomenclature at end of paper}} (length
  versus width) is made larger. In the formulas of this paper
  the aspect ratio $K$ appears as a free parameter. This
  allows us to e.g.~compare the packing efficiency of different sail concepts
  in a regime where both are configured with the same aspect ratio.}

The structure of the paper is as follows.  We first treat the
traditional photonic heliogyro and its issues, then consider the
improved FGPB heliogyro variant and then consider a scalable and
modular E-sail by adding tethers to the heliogyro. After that we
outline analogous LEO deorbiting applications for both E-sail and
heliogyro, discuss briefly FGPB implementation and demonstration
options and close the paper with a summary and
outlook. 

\section{Review of traditional heliogyro}
\label{sect:TradHeliog}

\begin{figure}
\centerline{\includegraphics[width=0.45\columnwidth]{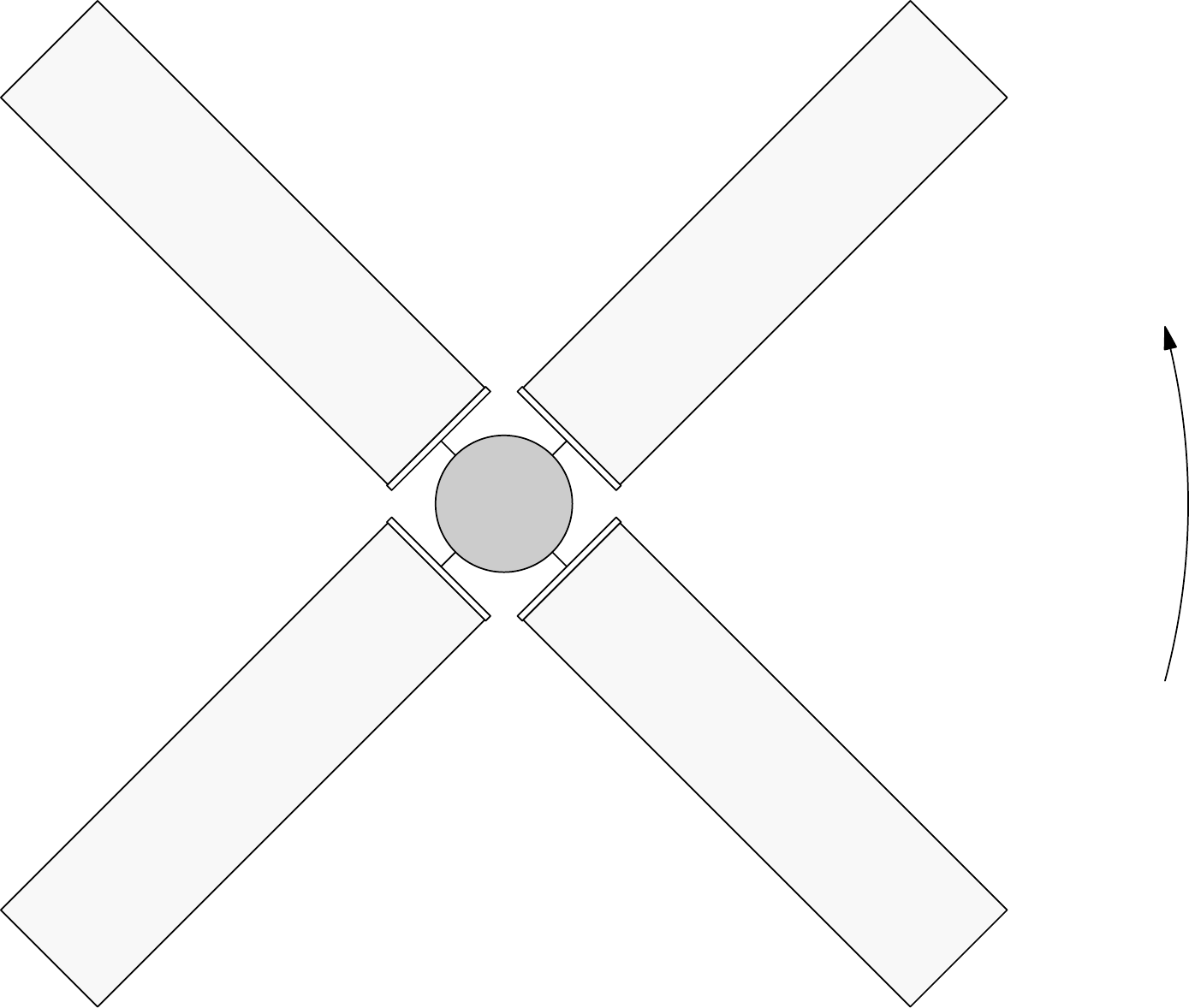}}
\caption{
Traditional \MARKII{spinning} heliogyro photon sail, in this case with four
blades. \MARKII{The blades are initially deployed from
wide rotating reels by the centrifugal force. Angular momentum
of the sail can be changed by photonic torque
resulting from blade tilting by mechanical
actuators on the main spacecraft. Said photonic torque can also be used to
turn the spin plane of the sail which enables control of the thrust
vector direction.}
}
\label{fig:TradHeliog}
\end{figure}

The traditional heliogyro \cite{MacNeal1967,Blomqvist2009}
(Fig.~\ref{fig:TradHeliog}) is a type of solar photon sail which has
several benefits compared to three-axis stabilised square
sails. \MARKIII{Recently the heliogyro concept has received renewed
interest with more detailed dynamical models of blade dynamics
\cite{GuerrantEtAl2012a} and news strategies for managing blade twisting \cite{GuerrantEtAl2012b}.} The
heliogyro spacecraft deploys its sheetlike photonic blades from
storage reels located on its perimeter. The blade tilting angle
(analogous to the angle of attack of helicopter blades) can then be
controlled mechanically from the main spacecraft to enable photonic
thrust vectoring with or without spinplane turning as well as to have
a capability to modify the spin rate \cite{Blomqvist2009}. The
heliogyro is simple to manufacture and to qualify because the sail
membrane is not a large continuous two-dimensional piece but instead
consists of rolled sheets similar to those normally used in industry
to store and transport thin film materials. Also, because the sail
material is nowhere folded and the system uses the centrifugal force
for deployment and stretching, the deployment sequence can be
demonstrated on ground simply by simulating the centrifugal
acceleration by Earth's gravity field. The Coriolis acceleration of
the spinning configuration in space cannot be simulated by gravity on
Earth, but it can be made small by selecting a slow speed of
deployment.

An issue with the heliogyro is, however, that with the limited physical
dimensions usually available for the main spacecraft inside the
payload fairing of the launch
vehicle, the total area of the photonic sail is limited. Consider the
case where the radius $R$ of the stowed configuration has some set value
and let us compute the maximal sail area as a function of the
number $N$ of the blades, under the assumption that the maximum aspect
ratio of the blade (length versus width) is $K$. To simplify the
analysis let us ignore the thickness of the reels. The sail is then
spanned by a regular polygon enclosed within a circle with radius $R$. The side length of the
polygon is $\MARKIII{h}=2R\sin(\pi/N)$, the blade length $L=K\MARKIII{h}$ by our
assumption, and the total area of the blades is given by
\begin{equation}
A = NL\MARKIII{h} = KN\MARKIII{h}^2 = 4KNR^2 \sin^2\left({\frac{\pi}{N}}\right).
\label{eq:A}
\end{equation}
Table \ref{table:heliogyro} gives the sail area $A$ for various $N$
from Eq.~(\ref{eq:A}). Also the area of the largest inscribed disk
inside the polygon $A_p=\pi R^2 \cos^2(\pi/N)$ is given which is
representative of the space available for the payload. We see from
Table \ref{table:heliogyro} that the optimum choice in terms of sail
area is $N=3$, but $N=4$ is almost as good and it provides a two times
larger payload space. We therefore adopt the four blade design
(Fig.~\ref{fig:TradHeliog}) as representative of the traditional
heliogyro in this paper. \MARKI{We emphasise that the four blade
  optimum applies only to the traditional heliogyro, not the new FGPB
  heliogyro which is introduced in the next section.}

\begin{table}[t]
\caption{Sail area and payload space of traditional heliogyro.}
\vskip2mm
\centering
\begin{tabular}{lll}
\hline
$N$ & Sail area & Payload space \\[0.cm]
\hline
\noalign{\vskip1mm}  
2 & $8\, K R^2$    & $0$               \\[0.cm]
3 & $9\, K R^2$    & $0.25  \pi R^2$   \\[0.cm]
4 & $8\, K R^2$    & $0.5   \pi R^2$   \\[0.cm]
5 & $6.91\, K R^2$ & $0.655 \pi R^2$   \\[0.cm]
6 & $6\, K R^2$    & $0.75  \pi R^2$   \\[0.cm]
7 & $5.27\, K R^2$ & $0.812 \pi R^2$   \\[0.cm]
8 & $4.69\, K R^2$ & $0.854 \pi R^2$   \\[0.cm]
\hline
\end{tabular}
\label{table:heliogyro}
\end{table}

The maximum value of $K$ is limited by a risk that the blades might no
longer be kept straight by the centrifugal force is they are
excessively long. A maximal $K$ of 1000 was reported in the early
analysis \cite{MacNeal1967} and more recent numerical calculations
were in agreement with this \cite{Blomqvist2009}. However, a more
conservative value of $K$ such as $K\approx 200$ might be preferable
in first flight models.

If $K=1000$ and $R=1.5$ m so that the spacecraft fits in a medium
class launch vehicle, the maximum sail area for $N=4$ is 18000 m$^2$,
providing about 140 mN of thrust at 1 au. The membrane mass is then
205 kg if made of similar 7.6 $\mu$m polyimide film as the IKAROS
solar sail \cite{IKAROS}. If the total mass of the spacecraft is taken
to be e.g.~500 kg, the characteristic acceleration is 0.2
mm/s$^2$ at 1 au.

\section{\MARKII{FGPB} heliogyro with freely guided blades}
\label{sect:NewHeliog}

\begin{figure}
\centerline{\includegraphics[width=0.45\columnwidth]{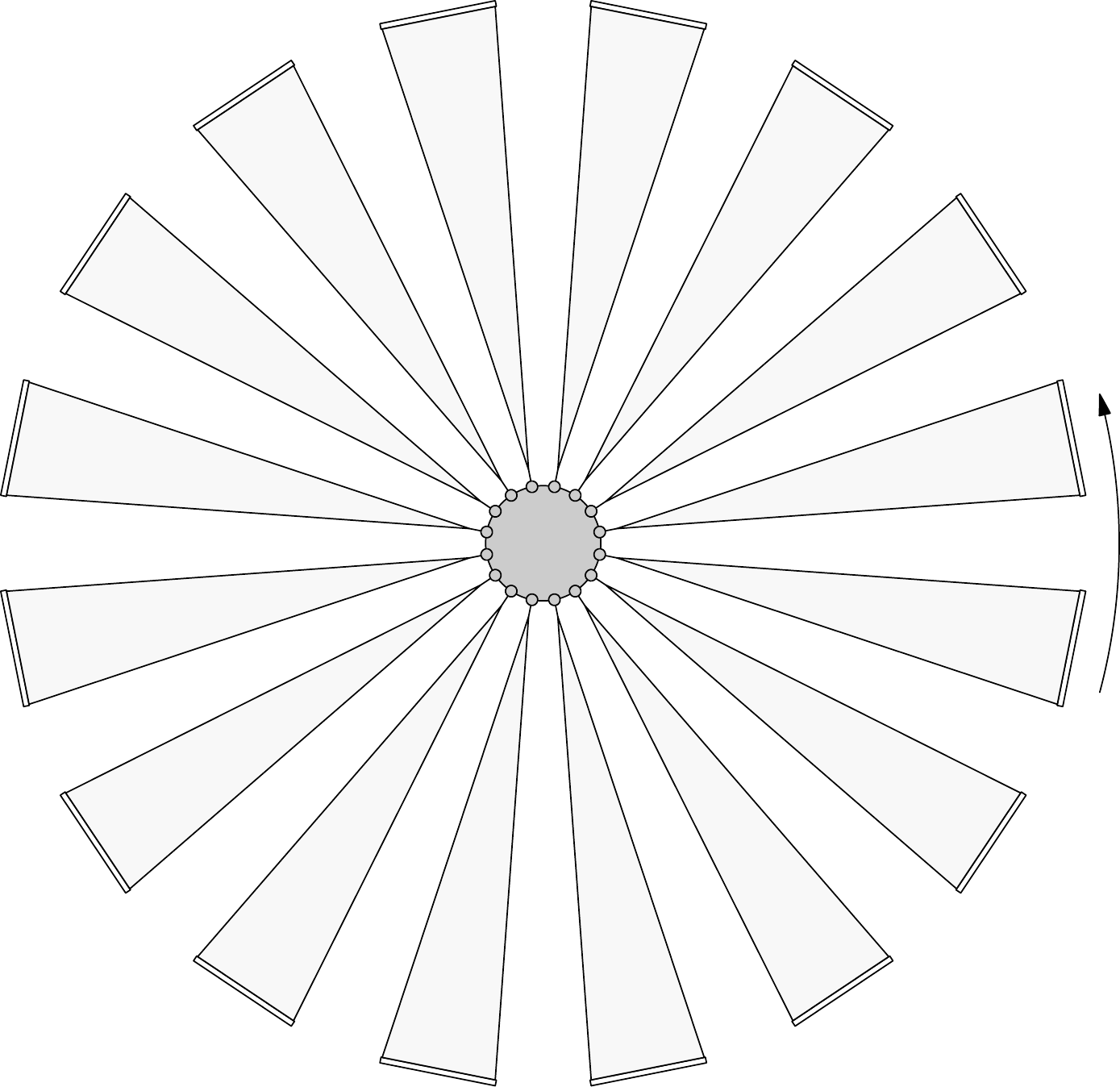}}
\caption{
\MARKII{Novel FGPB heliogyro concept. Blade attack angles are actuated
photonically by a moving ballast mass at the far end or by
modifiable blade optical properties.  For the same blade aspect
ratio and main spacecraft diameter, the displayed 16-blade sail has
two times larger sail area than the 4-blade traditional heliogyro of
Fig.~\ref{fig:TradHeliog}.}
}
\label{fig:FGPBHeliog}
\end{figure}

\begin{figure}
\centerline{\includegraphics[width=0.5\columnwidth]{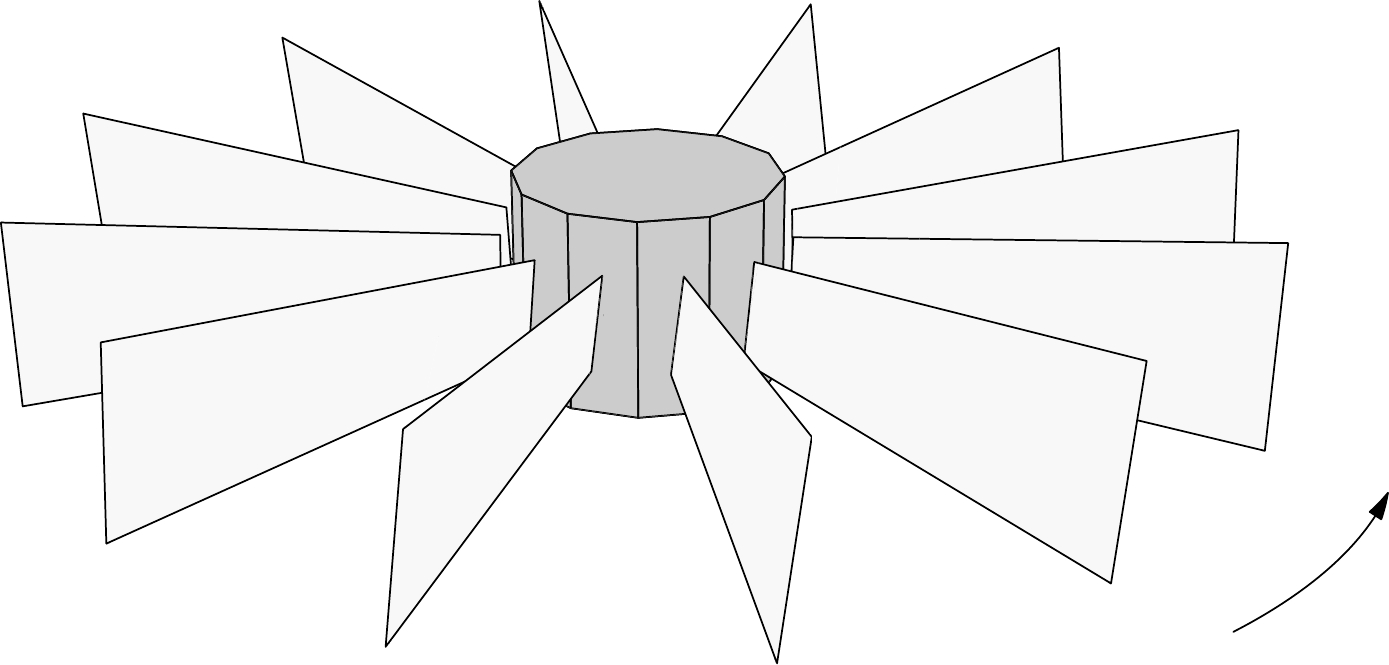}}
\caption{
\MARKII{The novel FGBP heliogyro concept in the early stage of deployment.}
}
\label{fig:DeployNear}
\end{figure}

\begin{figure}
\centerline{\includegraphics[width=0.5\columnwidth]{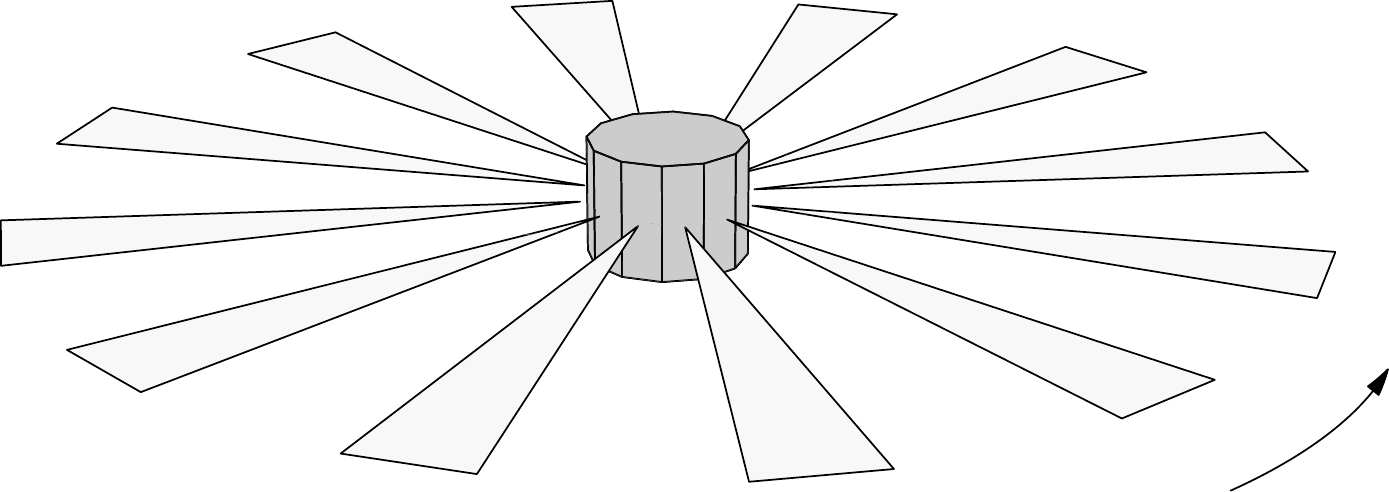}}
\caption{
\MARKII{The novel} FGBP heliogyro \MARKII{concept} after deployment.
}
\label{fig:DeployFar}
\end{figure}

Consider a heliogyro with triangular blades, the tipped ends of the
blades being attached to the spacecraft after deployment
(Figs.~\ref{fig:FGPBHeliog}-\ref{fig:DeployFar}). The deployment is
performed from rolls that are mounted almost vertically on the main
spacecraft. In the early phase of the deployment
(Fig.~\ref{fig:DeployNear}) the blades remain almost vertical because
the centrifugal force prevents them from twisting. At the end of the
deployment (Fig.~\ref{fig:DeployFar}) the tipped end of the blade is
however free to rotate and the blades align themselves horizontally.
This is so because mass elements that are not exactly on the blade's
longitudinal axis are slightly farther away from the spacecraft's axis
of rotation when the blade lies horizontally compared to when it is
oriented vertically.

\MARKII{In deployed configuration (Figs.~\ref{fig:FGPBHeliog} and
  \ref{fig:DeployFar}) the attack angle of the blades is controlled
  continuously by changing the blade's centre of mass relative to its
  centre of photon pressure. The actuation can be done by a moving
  mass at the end of the blade as sketched in Fig.~\ref{fig:Hang}; the
  gravity field shown in Fig.~\ref{fig:Hang} is however now replaced
  by the centrifugal acceleration which is due to the sail's
  spin. Alternatively, one could control the blade's attack angle by
  changing its optical properties so that the centre of photon
  pressure changes. Whereas the blade attack angles are controlled by
  mechanical twisting actuators from the main spacecraft in the
  traditional heliogyro (Fig.~\ref{fig:TradHeliog}), in the FGPB
  heliogyro they are controlled by the blade itself. This requires
  that the blades are autonomous units which are commanded from the
  main spacecraft, but the benefit is that more sail area can be
  fitted in a given launcher volume (Fig.~\ref{fig:FGPBHeliog} and
  subsection \ref{subsect:packingefficiency} below).}

\begin{figure}
\centerline{\includegraphics[width=0.5\columnwidth]{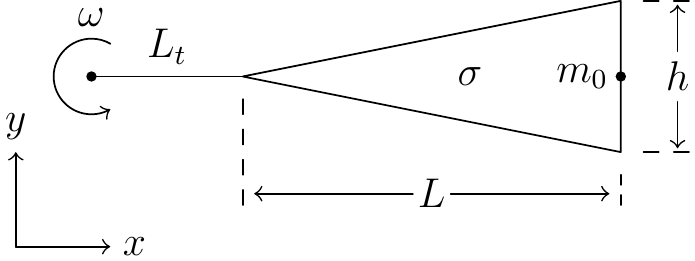}}
\caption{
Configuration for calculating centrifugal potential energy and
twisting torque.
}
\label{fig:torquecalc}
\end{figure}

\MARKIII{As mentioned above, the heliogyro has several theoretical
  advantages relative to other photon sails concepts such as no need
  for long rigid booms, no need to fold the sail and the possibility
  of using roll to roll methods in manufacturing of the sail. On the other
  hand, a heliogyro platform must always spin which can be an
  inconvenience for missions requiring accurate pointing or docking.}

\subsection{\MARKII{Calculation of flattening natural torque}}

Let us now calculate the torque that tends to make the blade to lie
horizontally in the spin plane. Consider a simplified model of the
triangular blade in Fig.~\ref{fig:torquecalc}. We assume that the
blade (height \MARKII{$h$}) has uniform areal mass density
\MARKII{$\sigma$} and an additional point mass $m_0$ resides at the
middle of the blade's far end, representing the mass of the remote
unit needed to actuate the centre of mass change. We calculate the
change in the centrifugal potential energy when the blade is tilted to
an angle angle $\alpha$ from the spin plane. a nonzero $\alpha$
situation is similar to the geometry shown in
Fig.~\ref{fig:torquecalc} except that the blade's height $h$ becomes
less and its areal mass density $\sigma$ is correspondingly increased:
\begin{eqnarray}
\MARKII{h'}       &=& \MARKII{h} \cos\alpha \nonumber \\
\MARKII{\sigma'}  &=& \frac{\MARKII{\sigma}}{\cos\alpha}.
\end{eqnarray}

The centrifugal potential energy of the blade is
\begin{eqnarray}
V &=& -\frac{1}{2}\omega^2 \int dm r^2 \nonumber\\
&=& -\omega^2 \MARKII{\sigma'} \int_{\tetherlen}^{\tetherlen+L} dx
\int_0^{\left(\frac{x-\tetherlen}{L}\right)\frac{\MARKII{h'}}{2}} dy (x^2+y^2) \nonumber\\
&=& \frac{-\omega^2 \MARKII{\sigma' h'}}{2}\left[
\int_{\tetherlen}^{\tetherlen+L}dx x^2 \left(\frac{x-\tetherlen}{L}\right)
+ \frac{1}{\MARKII{12}}\int_{\tetherlen}^{\tetherlen+L}dx\left(\frac{x-\tetherlen}{L}\right)^3 \MARKII{h'}^2
\right].
\end{eqnarray}
For calculating the twisting torque, only terms that depend on
$\alpha$ are needed. The product \MARKII{$\sigma' h'$} is equal to
\MARKII{$\sigma h$} so it does not depend on $\alpha$. Thus, only the
last integral term depends on $\alpha$ through \MARKII{$h'^2$} and we obtain
\begin{eqnarray}
V &=& {\rm const}
- \omega^2 \MARKII{\sigma h} \frac{1}{24} \int_0^L dx \frac{x^3 \MARKII{h'}^2}{L^3}
\nonumber\\
&=& {\rm const} - {1\over 96} \omega^2 \MARKII{\sigma h^3} L \cos^2\alpha
\nonumber\\
&=& {\rm const} + {1\over 96} \omega^2 \MARKII{\sigma h^3} L \sin^2\alpha
\end{eqnarray}
\MARKII{where in the last step we used $\cos^2\alpha=1-\sin^2\alpha$
and absorbed the offset $1$ into the unimportant constant.} 
The torsional torque is
\begin{equation}
\tau = -\frac{dV}{d\alpha}
= -{1\over 96} \omega^2 \MARKII{\sigma h^3} L \sin2\alpha
= -{1\over 48} \omega^2 m_{\rm b} \MARKII{h^2} \sin2\alpha
\label{eq:tau}
\end{equation}
where $m_b=(1/2)\MARKII{\sigma h} L$ is the mass of the triangular
blade. The angular equation of motion of the blade which describes
tilting oscillations is
\begin{equation}
\ddot{\alpha} = \frac{\tau}{I}
\end{equation}
where $I=(1/24)m_b \MARKII{h}^2$ is the blade's moment of inertia about its
longitudinal axis. Thus the \MARKI{angular} equation of motion becomes
\begin{equation}
\ddot{\alpha} = -\frac{1}{2} \omega^2 \sin2\alpha
\end{equation}
which for small oscillations $\alpha\approx 0$ reduces to
$\ddot{\alpha}=-\omega^2\alpha$. Because this is the equation of a
harmonic oscillator at frequency $\omega$, we conclude that \MARKIII{in the limit of small oscillations i.e. linear dynamics}, the
oscillation angular frequency of small tilting oscillations is the
same as the sail's spin period. A small extra calculation shows that
this result is valid also for other than triangular shaped blades.

The centrifugal potential energy of the mass $m_0$ did not enter the
calculation because its distance from the spin axis does not depend on
$\alpha$. Of course, once the mass is actively moved from the symmetry
position as in Fig.~\ref{fig:Hang}, the value of $m_0$ affects on how
fast the blade's orientation responds to the actuation.

We must now consider if the natural torque given by Eq.~(\ref{eq:tau})
which tends to zero the blade's tilting angle $\alpha$ is typically
much smaller than a conveniently achievable actuating torque caused by
shifting the centre of mass with respect to the centre of the photon
pressure, for values of $\alpha$ that one wants to
achieve. \MARKII{The worst case (largest natural torque) occurs when the tether length
  $L_t=0$. In this case} the
centrifugal force acting on the blade is (\MARKII{$L=Kh$} is the blade length)
\begin{eqnarray}
F_{\rm cf}^{\rm b} &=& \int_0^L dr h\left(\frac{r}{L}\right)\sigma r \omega^2\nonumber\\
&=& \sigma \omega^2 \left(\frac{h}{L}\right)\int_0^L dr r^2\nonumber\\
&=& \frac{1}{3} \sigma \omega^2 K^2 h^3.
\label{eq:Fcf}
\end{eqnarray}
The total centrifugal force $F_{\rm cf}$ is the sum of $F_{\rm cf}^{\rm b}$ and
a centrifugal force acting on the $m_0$-mass remote unit, $F_{\rm
  cf}^{\rm RU}=m_0 Kh \omega^2$.

The photonic thrust exerted on the triangular blade is given by
\begin{equation}
F = \frac{1}{2} K \MARKII{h'}^2 P_{\rm rad}
\end{equation}
where $P_{\rm rad}$ is the solar radiation pressure. We demand that $F
\ll F_{\rm cf}$ so that blade bending due to radiation pressure is small. We
parametrise this by writing $F_{\rm cf}=k F$ where $k$ is a large
enough number, e.g.~$k=10$.
We then compute by which distance $y$ the centre of mass has to differ
from the centre of radiation pressure of the blade to balance the
natural torque $\tau$ (Eq.~\ref{eq:tau}) by an actuated photonic
torque $\tau'$ which is
(as an order of magnitude estimate ignoring some geometrical factors)
given by $\tau'=F y$. From the torque balance requirement \MARKII{$\tau'=\tau$} we solve
\MARKII{for $y$. Using Eqs.~(\ref{eq:tau}), (\ref{eq:Fcf}) and $L=Kh$ we obtain}
\MARKII{
\begin{eqnarray}
y &=& \tau/F
= k \tau/F_{\rm cf} \nonumber\\
&\le& k \tau/F_{\rm cf}^{\rm b} \nonumber\\
&=& k \frac{3}{96}
\frac{\omega^2 \sigma h^3 L \sin 2\alpha}
{\sigma \omega^2 K^2 h^3} \nonumber\\
&=& \frac{1}{32} \left(\frac{k}{K}\right)h \sin 2\alpha.
\end{eqnarray}
}

Inserting reasonable conservative values $k\approx 10$ and $K\approx
200$ one sees that the resulting $y$ is nearly three orders of
magnitude smaller than the width of the blade $h$. Including the
effect of the remote unit $m_0$ would make $y$ even smaller because
for a given blade tension it would decrease $\omega$ and thus the
natural torque $\tau \sim \omega^2$. Thus, the actuating photonic
torque can easily overcome the natural restoring torque.

\begin{figure}
\centerline{\includegraphics[width=0.5\columnwidth]{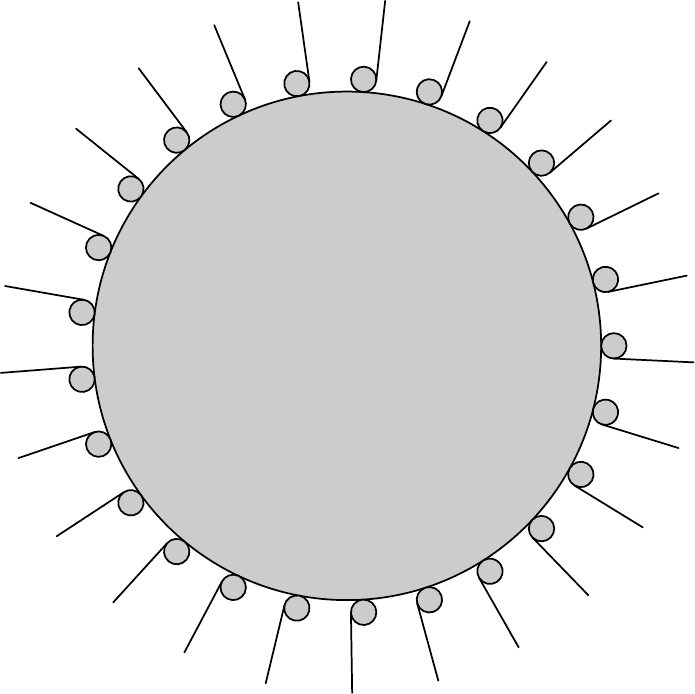}}
\caption{
Top view of \MARKII{the FGPB} heliogyro with assumed cylindrical
spacecraft. The reels from which the blades are deployed resides along
the perimeter.
}
\label{fig:TopView}
\end{figure}

\subsection{\MARKII{Packing efficiency of FGPB heliogyro}}
\label{subsect:packingefficiency}

Figure \ref{fig:TopView} shows a top view of the \MARKII{FGPB} heliogyro
design in case of a cylindrical main spacecraft. In reality the reels
would not be installed exactly vertically, to give the blades a small
initial obliqueness so that they would settle in a predictable
direction without a need for photonic actuation during deployment.

We shall now estimate how much total sail area can be packed in a
cylindrical spacecraft of radius $R$. We assume that the height of the
cylinder is $h=R \sqrt{3}$, because then all three principal moments of
inertia coincide if the mass density of the cylinder is
homogeneous. \MARKII{In this way} we maximise the height of the cylinder to maximise the
blade area \MARKII{while keeping} the spin-aligned inertial moment
(marginally) larger than the other components to ensure
\MARKII{rotational stability} in all phases.  We also assume that the reels are positioned
along the cylinder's wall by leaving an equal amount of clearance
between them than the reel diameter. This is to leave room between the
reels where to store the remote units. \MARKII{Then the number of
blade reels with radius $r_{\rm R}$ that fit on the perimeter of the
cylinder $2\pi R$ is (Fig.~\ref{fig:TopView})}
\begin{equation}
N \MARKII{= \frac{2\pi R}{4 r_{\rm R}}} = \frac{\pi}{2} \frac{R}{r_{\rm R}}.
\end{equation}
\MARKII{If one neglects the inner radius of the reel, the storage
volume of a blade reel is $\pi r_{\rm R}^2 h$ which must be equal to
the stowed volume of the blade $Lhd=Kh^2 d$ where $d$ is the blade's
thickness. From this condition one can solve the reel radius $r_{\rm R}$ to obtain}
\begin{equation}
r_{\rm R} = \sqrt{\frac{Khd}{\pi}}.
\end{equation}
The total area of the blades is
\begin{equation}
A^{\rm new} = \frac{1}{2}NKh^2
= \frac{\left(\pi\sqrt{3}\right)^{3/2}}{4} \sqrt{\frac{K}{d}} R^{5/2}
= 3.17 \sqrt{\frac{KR}{d}} R^2.
\label{eq:newA}
\end{equation}

Equation (\ref{eq:newA}) can be compared with the corresponding law of
the traditional $N=4$ heliogyro for which $A = 8 K R^2$ (Table
\ref{table:heliogyro}). Their ratio is
\begin{equation}
\frac{A^{\rm new}}{A^{\rm old}} = 0.40 \sqrt{\frac{R}{d}\frac{1}{K}}.
\end{equation}
The factor $R/d$ is large whereas $K$ is at most 1000. If $R=1$ m,
$d=7.6\,\mu$m and $K=1000$, the area ratio is $A^{\rm new}/A^{\rm
  old}=4.6$. Thus the \MARKII{FGPB} heliogyro concept allows one to
pack more sail area in a spacecraft of a given
cylindrical volume. If one uses a more conservative blade aspect ratio
such as $K=200$, the relative benefit of the \MARKII{FGPB} design becomes
larger. Also, if one introduces a thinner sail material (smaller $d$),
the \MARKII{FGPB} design can host a larger sail without increasing $R$, whereas
for the traditional heliogyro the maximum area of the sail is
essentially independent of the membrane thickness.

\section{E-sail with FGPB spin management}
\label{sect:Esail}

\begin{figure}
\centerline{\includegraphics[width=0.5\columnwidth]{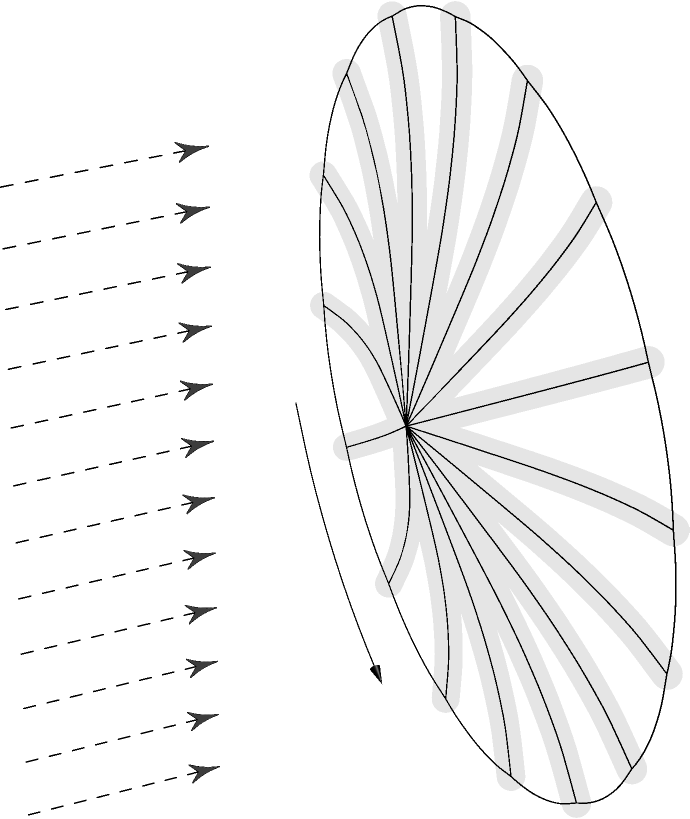}}
\caption{
Traditional E-sail design where auxiliary tethers that connect the
tips of the main tethers together ensure dynamical stability.
}
\label{fig:AuxtetherEsail}
\end{figure}

The traditional E-sail (Fig.~\ref{fig:AuxtetherEsail}) uses auxiliary
tethers to ensure that the main tether do not collide in the varying
solar wind \cite{RSIpaper}. The remote units at the tips of the main
tethers need some spinup thruster and the same thruster can also be
used to modify the spin rate during E-sail flight. The thruster can be
a cold gas thruster, a FEEP thruster \cite{MarcuccioEtAl2009}, a
photonic blade thruster \cite{paper16} or some other thruster. The
photonic blade thruster as is the subject of this paper has the
benefit of having infinite specific impulse so that there the
capability to modify the spin rate during the mission is
large. Modification of the spin rate may be needed to overcome a
secular change of the spin rate due to the heliocentric orbital
Coriolis force \cite{paper14,paper16}. In specific missions one might
also want to modify the spin rate for some other reason.

\begin{figure}
\centerline{\includegraphics[width=0.5\columnwidth]{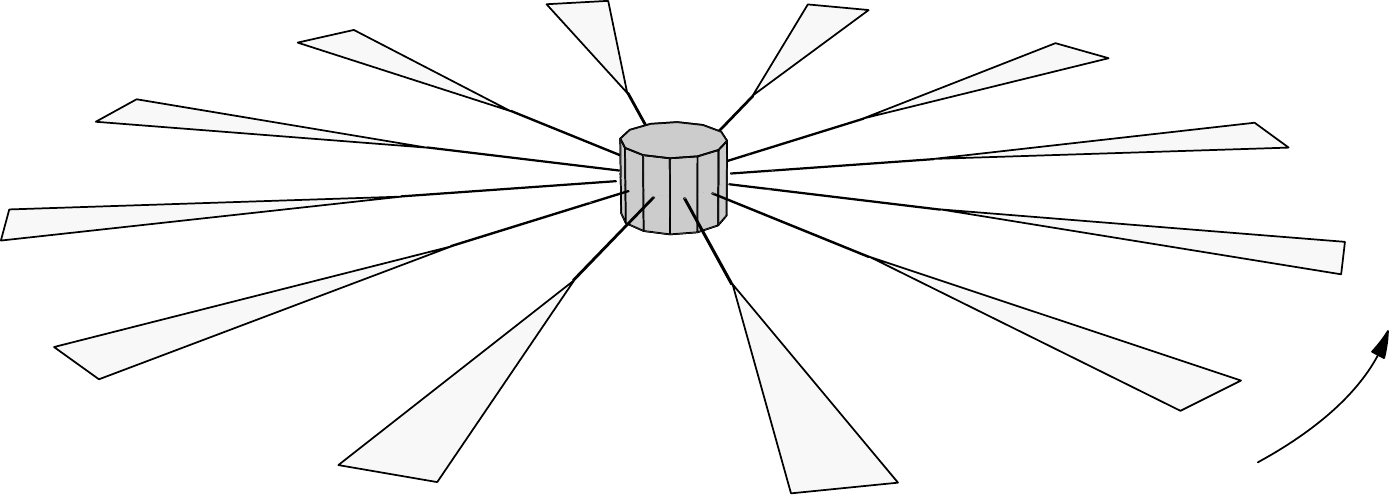}}
\caption{
E-sail with freely guided photonic blades spin management. In reality
the tethers would be much longer than depicted here.
}
\label{fig:DeployTeth}
\end{figure}

We also introduced the idea \cite{paper16} that by making the photonic
blades larger one could get rid of the auxiliary tethers which would
simplify the E-sail design and make it more scalable and modular
(Fig.~\ref{fig:DeployTeth}).  In this case the photonic thrust must be
large enough to overcome the tendency of the tethers to come together
by random changes of the solar wind.  This variant of the E-sail is
scalable and modular because the number and length of the main tethers
are no longer tied together by a necessity to connect their tips by
auxiliary tethers. This configuration is entirely similar to the FGPB
heliogyro considered in section \ref{sect:NewHeliog}. The only difference is
the existence of a long E-sail tether between the main spacecraft and
the photonic blade, and the fact that the photonic blade is smaller
because it is only used as an auxiliary propulsion device for the
E-sail control purpose.

In the traditional E-sail which uses auxiliary tethers, if the tether
length is 20 km, the E-sail force per length 500 nN/m their and the
tether tension 5 cN, we found earlier \cite{paper16} that 2.7 m$^2$
blade area per tether is sufficient and that the blade area
requirement is independent of the heliocentric distance. For the new
design which does not have auxiliary tethers, we made some numerical
experiments where the E-sail is flown in real solar wind that for the
nominal 20 km E-sail tether length, the required blade area per tether
is 16-20 m$^2$.

In the \MARKII{FGPB} design one could have, for example, a low-end E-sail with
four tethers, each 20 km long, and 40 mN total nominal E-sail thrust
at 1 au. The total mass of the blade membranes would be 1.5 kg, if
each is 20 m$^2$ large and made of 12.6 $\mu$m ITAR-free thin metallised kapton
while the E-sail tethers would weigh 0.9 kg if made using our standard
4-fold Heytether construction with 25 $\mu$m loop and 50 $\mu$m base
aluminium wire \cite{SeppanenEtAl2011}. Allowing for 0.3 kg for each
main tether reel, 0.3 kg for each remote unit which controls the
blade's orientation by shifting the centre of mass or other method
\cite{paper16}, and 1.2 kg for the 40 W/20 kV electron gun and its
high voltage source, the total mass of the E-sail hardware would be
only 6 kg in this case, yielding a specific acceleration at 1 au of
6.7 mm/s$^2$. By specific acceleration we mean the thrust divided by
the propulsion system's (initial) mass. For example, this 6 kg E-sail
device could carry a spacecraft whose total mass is 80 kg at 0.5
mm/s$^2$ characteristic 1 au acceleration which is equivalent to 15
km/s/year delta-v capability. Furthermore the E-sail acceleration
decays as $1/r$ with solar distance $r$ which is slower than the
$1/r^2$ effective thrust decay of photonic sail and solar electric
propulsion.

We also want to point out that there is a natural continuum of hybrid
E-sail/photonic sail engineering possibilities between the purely
photonic heliogyro and the photonically guided E-sail. These hybrid
designs would use the E-sail effect in interplanetary space and revert
to photonic propulsion inside the magnetosphere where the solar wind
does not exist. In hybrid design the photonic blades would typically
be much larger in area than what is needed for the E-sail control
alone.

\section{Deorbiting applications of FGPBs}

In low Earth orbit (LEO), force due to the atmospheric molecular flow
overcomes the solar photonic force typically below 800 km
altitude. Also in LEO, charged tethers can be used for generating a
Coulomb drag plasma brake effect \cite{paper3,paper8}. The plasma
brake is similar to the E-sail except that it uses a negatively biased
tether to keep the gathered current and power consumption small. It
seems that a negative tether is more beneficial in the ionosphere
while a positive tether is better in the solar wind
\cite{paper3,RSIpaper}. Both positive and negative polarity tethers
generate a Coulomb drag effect where force is exerted on the tether
which is aligned with the tether-perpendicular component of the plasma
flow in the tether's frame of reference.

One could use the FGPB heliogyro device of section
\ref{sect:NewHeliog} (Fig.~\ref{fig:DeployFar}) in LEO directly as a
neutral drag deorbiting device. The photonic pressure would simply be
replaced by or augmented by the pressure due to the molecular ram flow
of the upper atmosphere. Manoeuvring remains at least qualitatively
similar because molecular flow behaves much in a same (although not
completely similar) way as the radiation pressure.

Analogously, a device similar to the FGPB E-sail described in section
\ref{sect:Esail} (Fig.~\ref{fig:DeployTeth}) could be tuned for LEO
deorbiting. The tether lengths and thickness might have to be changed
to accommodate a possibly larger Coulomb drag (depending on the plasma
density and thus altitude) and to resist the tendency of Earth's
gravity gradient to disturb a spinning tether rig. Above $\sim 800$ km
the blades would use solar photonic pressure for tether dynamical
control (similar to the E-sail), below this altitude they would
primarily use the atmospheric drag for the same purpose. If wanted for
some mission specific reasons, the traditional electrodynamic tether
Lorentz force could also be available from the same tethers by
charging them positively and thereby letting them collect a larger
current. The active electrodynamic tether mode would allow also orbit
inclination change and orbit raising manoeuvres in addition to only
braking. The original single-tether gravity stabilised plasma brake
device \cite{paper8} can deorbit up to $\sim 100$ kg debris payloads
(an exact mass limit cannot be given since it depends on the time
available for the deorbiting task). With a spinning tether rig, the
device could be scaled up by the number of tethers, thus making it
potentially feasible to apply the efficient plasma brake effect to
multi-tonne debris object orbit lowering tasks. Of course, for large
debris objects which do not burn completely in the atmosphere, one
would still have to use a chemical kick in the final phase so that the
object reenters over an uninhabited ocean area.

\section{Implementation and demonstration of FGPB}

Implementing the freely guided photonic blade calls for a controllable
way of creating a difference between the centre of mass and centre of
photonic pressure of the blade. One way to do it is to have a moving
mass as shown in Fig.\ref{fig:Hang}. In our earlier paper we similarly
considered a bar-shaped remote unit inside which there is a sliding
mass \cite{paper16}.

\begin{figure}
\centerline{\includegraphics[width=0.7\columnwidth]{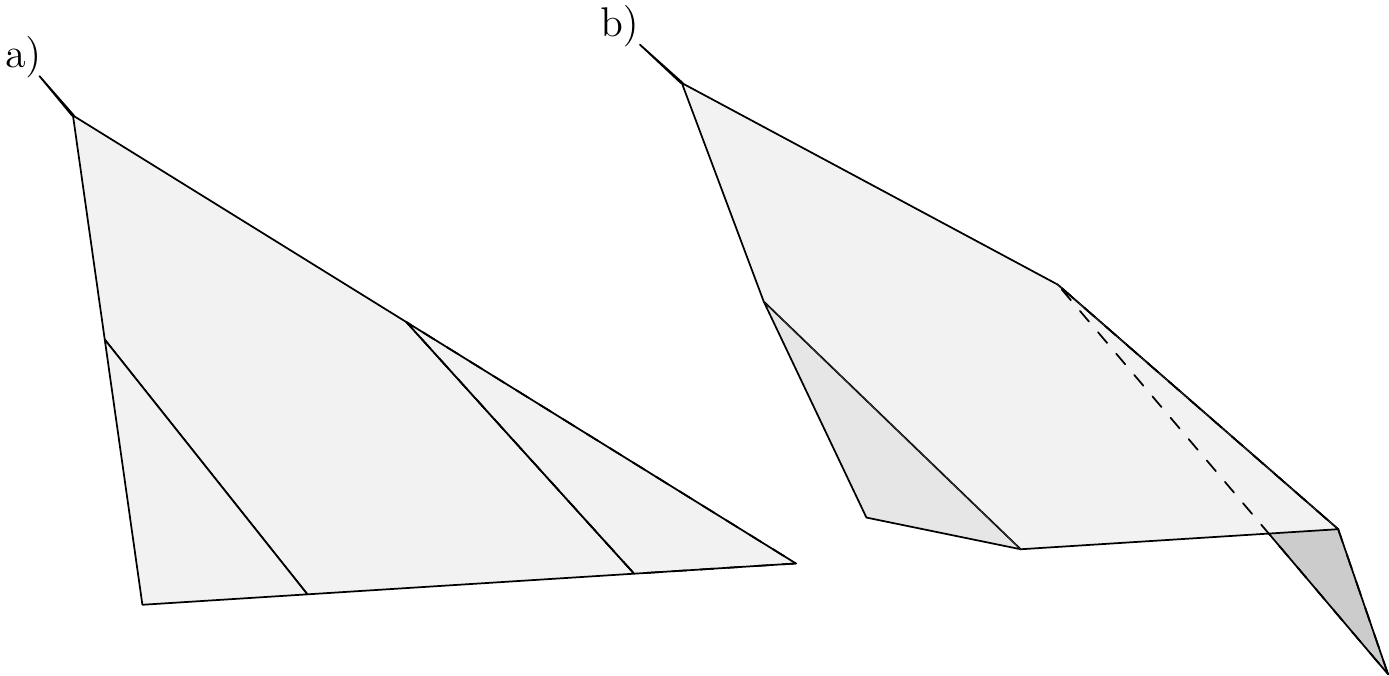}}
\caption{
Flap actuated FGPB in (a) unactuated and (b) exemplary
actuated position.
}
\label{fig:FoldBlade}
\end{figure}

If flat bending devices are available, the FGPB might also be actuated
by flaps as shown in Fig.~\ref{fig:FoldBlade}. In this case, even the
control electronics might be printed on the blade so that no physical
``remote unit'' would be needed since the blade itself would contain
the flap actuated mechanism. Alternatively, if one wants to avoid
bending parts, actuation could be accomplished by modifying the
optical properties of the blade. The Japanese IKAROS mission uses
technology borrowed from liquid crystal displays (LCDs) for this
purpose \cite{IKAROS}. One might also cover a significant fraction of the blade by
thin film solar panels and modify their optical properties by
short-circuiting them. The power gathered by the non short circuited
areas could be used for resistive heating of a suitable near edge area
of the blade to enhance the photonic torque available from such
actuation.

To increase the technical readiness level (TRL) of the FGPB to 4-5,
the controllability of the FGPB should be demonstrated on ground in a
vacuum chamber equipped with a solar simulator or other artificial
photon source, by using Earth's gravity field to simulate the
centrifugal force (Fig.~\ref{fig:Hang}). One would demonstrate a
capability to turn the hanging photonic blade to an angle given by
remote command by a controlled change of its centre of mass or optical
properties.

To increase the TRL of FGPB to 7, the system could be taken to a
CubeSat for orbital demonstration. One would set the CubeSat to spin
e.g.~by magnetorquers and then deploy the blade from a roll followed
by a tether, similar to the ESTCube-1 CubeSat which is launched in
March 2013 to measure the E-sail effect in orbit
\cite{RSIpaper,PajusaluEtAl2012}. One would then demonstrate that one
can tilt the blade controllably by photonic thrust and thereby modify
the spin state of the tether plus blade system according to
commanding. One could use 1-4 FGPB tethers depending on how much space
there is available on the particular CubeSat. One tether is enough to
demonstrate the necessary controllability and manoeuvres, however,
because multiple tethers are dynamically only weakly dependent on each
other through the central spacecraft. After reaching TRL 7 with one
tether, one could probably build multi-tether FGPB E-sails and
heliogyros for interplanetary travel and deorbiting without additional
system-level space demonstrations.

\section{Summary and outlook}

\begin{table}
\caption{\MARKII{Photonic sail concepts.}}
\vskip2mm
\centering
\begin{tabular}{ll}
\hline
\MARKII{Concept} & \MARKII{Characteristics} \\
\hline
& \\[-4.5mm]
\MARKII{Square sail} & \MARKII{+High TRL 7 in small scale}\\
& \MARKII{-Scaling is nontrivial problem} \\
\MARKII{Traditional heliogyro} & \MARKII{-Scalability limited by launcher diameter} \\
\MARKII{FGPB heliogyro} & \MARKII{+Scalable} \\
& \MARKII{-Needs continuous active control} \\
\hline
\end{tabular}
\label{table:photonic}
\end{table}

\begin{table}
\caption{\MARKII{Electric solar wind sail concepts.}}
\vskip2mm
\centering
\begin{tabular}{ll}
\hline
\MARKII{Concept} & \MARKII{Characteristics} \\
\hline
\MARKII{E-sail with aux.~tethers} & \MARKII{+Reasonably high TRL 4-5}\\
& \MARKII{-Auxiliary tether ring is single point failure}\\
\MARKII{FGPB E-sail} & \MARKII{+Very good scalability and modularity}\\
& \MARKII{+No single point failures}\\
& \MARKII{-Needs continuous active control} \\
\hline
\end{tabular}
\label{table:electric}
\end{table}

\MARKII{
We surveyed propellantless propulsion applications of freely
guided photonic blades (FGPBs) in the context of photonic sails (Table
\ref{table:photonic}) and electric sails (Table
\ref{table:electric}). The FGPBs can be used as scalable photonic
sails directly or as control propulsion for E-sails. Hybrid photonic-E-sails
are also conceivable. Outside of Earth's
magnetosphere where the solar wind is available as a thrust source,
E-sails can provide potentially much higher performance than photonic
sails \cite{RSIpaper}.
}

We also showed that the blades have a
natural dynamical tendency to settle in the spin plane which is
actually helpful in deployment phase, and that the tendency
can be easily overcome by actuated photonic thrust during
flight. We identified and \MARKII{considered} the following \MARKII{specific} FGPB application domains:
\begin{enumerate}
\item \MARKII{FGPB} photonic heliogyro sail \MARKII{concept} whose packing density
  is higher than that of a traditional heliogyro.
\item Control propulsion for a scalable and modular variant of the E-sail which does not
  need auxiliary tethers for dynamical stability.
\item Hybrid photonic-E-sail devices.
\item Various deorbiting devices based on plasma brake and neutral
  drag effects. These devices are analogous to FGPB E-sails and heliogyros.
\end{enumerate}
We also briefly discussed engineering alternatives for FGPB
implementation (based on changing the centre of mass or changing the
blade's optical properties) and outlined a path how the FGPB's TRL
could be raised to 4-5 with simple ground-based demonstration and to 7
by a CubeSat demonstration.

We conclude that FGPB seems to be a technical concept that would enable 
a potentially large class of very significant applications in
propellantless interplanetary propulsion and deorbiting. \MARKII{FGPBs
  need continuous active control, but this drawback is probably not
  severe in comparison to the benefits.} We consider
that raising its TRL as well as looking in more details into its
application would therefore be highly motivated.

\section{Acknowledgement}


The research leading to these results has received funding from the
European Community's Seventh Framework Programme ([FP7/2007-2013])
under grant agreement number 262749. We also acknowledge the Academy
of Finland and the Magnus Ehrnrooth Foundation for financial support.


\appendix

\MARKII{\section{Nomenclature}}

\MARKII{

\begin{tabular}{ll}
$A$ & Heliogyro total blade area \\
$A_p$ & Area of largest inscribed disk inside polygon \\
$d$ & Blade thickness \\
$F$ & Photonic thrust force (acting on single blade) \\
$F_{\rm cf}$ & Centrifugal force (acting on single blade) \\
$h,h'$ & Blade \MARKIII{width (also called height)}, plain and projected \\
$I$ & Inertial moment (about long axis) of blade \\
$K$ & Heliogyro blade aspect ratio (length per width) \\
$k$ & $=F_{\rm cf}/F$, centrifugal force per photonic force\\
$L$ & Blade length \\
$L_t$ & Tether length \\
$m_0$ & Remote unit mass of novel type heliogyro blade \\
$m_b$ & Mass of single novel type heliogyro blade \\
$N$ & Number of blades or tether-blade systems \\
$P_{\rm rad}$ & Radiation pressure \\
$R$ & Radius of stowed configuration or spacecraft \\
$r$ & Radial distance (from spacecraft or from sun) \\
$r_{\rm R}$ & Blade storage reel outer radius \\
$V$ & Centrifugal potential energy of blade \\
$x$ & Horizontal coordinate \\
$y$ & Vertical coordinate, actuated shifting of centre of mass of blade\\
$\alpha$ & Tilting angle of novel type heliogyro blade \\
$\omega$ & Angular rotation rate of novel type heliogyro \\
$\sigma,\sigma'$ & Mass per area of blade, plain and projected \\
$\tau,\tau'$ & Torsional torque, plain and actuated \\
\end{tabular}
\vskip\baselineskip

\begin{tabular}{ll}
E-sail & Electric solar wind sail, electric sail \\
FEEP & Field effect (or field emission) electric propulsion \\
FGPB & Freely guided photonic blade \\
ITAR & International traffic in arms regulations, set of U.S.~regulations\\
LCD & Liquid crystal display \\
LEO & Low Earth orbit \\
RU & Remote unit, autonomous device at tip of each E-sail tether \\
TRL & Technical readiness level \\
\end{tabular}

}









\end{document}